\documentclass{elsart}

\usepackage[dvips]{graphicx}

\usepackage{amssymb}

\begin{document}

\begin{frontmatter}



\title{Properties of low variability periods in financial time series}
\author[ioc]{Robert Kitt\corauthref{cor1}}
\ead{kitt@ioc.ee}
\author[ioc]{Jaan Kalda},
\corauth[cor1]{Corresponding author. Tel.: +372 6204174; fax: +372 6204151}
\address[ioc]{Department of Mechanics and Applied Mathematics, Institute of Cybernetics at Tallinn University of Technology, 12061, Tallinn, ESTONIA}


\begin{abstract}
Properties of low-variability periods in the time series are analysed. The theoretical approach is used to show the relationship between the multi-scaling of low-variability periods and multi-affinity of the time series. It is shown that this technically simple method is capable of reveling 
more details about time series than the traditional multi-affine analysis. 
We have applied this scaling analysis to financial time series: a number of daily currency and stock index time series. The results show a good scaling behaviour for different model parameters. 
The analysis of high-frequency USD-EUR exchange rate data confirmed the theoretical expectations.
\end{abstract}

\end{frontmatter}

\section{Introduction}
\label{intro}
Perhaps the only thing which is more complicated and unpredictable than a human mind, is the collective human mind. This
collective human mind is the driving force of financial market fluctuations. The intrinsic complexity of the market
dynamics forms the basis for the exponential growth of the econophysics, since the publication of seminal papers of
Stanley et al \cite{stan1,stan2,stan3}. As in every new scientific discipline, the Econophysics is developing rapidly in
various directions. Research in those various fields has contributed several stylized statistical properties of asset
returns (which, in principle, may cease to be valid at a certain moment of time, because the collective human mind is
capable of learning), c.f.\ \cite{cont,lux1,lux2}. Most advancements in econophysics are owing to the concepts of 
scale-invariance, intermittency and heteroscedasticity. So, viable approximations for the market movements 
have been either based on the continuous time random walk (CTRW) model \cite{Montroll,sc1,sc2,sc3,sc4,sc5},
or derived from it, c.f.\ the model of fractional Brownian motion (fBm) in multifractal 
trading time (fBm-mtt) \cite{mandel3,mandel1,mandel2}.  
There is strong empirical evidence that fluctuations in financial markets possess multifractal statistics \cite{vande,ausl1,schmitt1,schmitt2,ausl2}. Multifractality has been found to be inherent both to the currency markets \cite{vande,schmitt1,schmitt2}, and to the stock exchange markets \cite{ausl2}. The time series analysis is usually based on the closing prices of two consecutive trading days, but higher frequency data have been also used \cite{taka1}.

The focus of this paper is placed on the multi-scaling properties of currency and stock exchange time series.
So, we refrain ourselves from disussing the advancements in understanding the other aspects of financial time series.

It has been recently shown that in the case of intermittent time series, the scaling behaviour of low-variability
periods can provide additional information, as compared to a multifractal analysis \cite{kalda2}. Preliminary study
of the time series of stock indices and currency rates indicated that similarly to the 
heart rate variability data, the financial time series are also characterized
by a multi-scaling behaviour of low-variability (i.e.\ ``silent'' or ``calm'') periods \cite{Kitt}. Such a behaviour has
been independently verified by Kaizoji et al \cite{kaz}.

In this paper, analytic approach is used to derive the relationship between the multi-scaling of low-variability periods,
and multi-affinity of the time series. Besides, a detailed analysis of the statistics of low variability periods is
performed for various daily stock indices and daily currency rates. The study of the high-frequency data of 
the USDEUR exchange rate indicates that at the time-scale of one day, 
there is a cross-over between two scaling laws for the low-variability periods.

\section{Theoretical approach to low-variability periods}
\label{theory}
Most modern studies of the variability of intermittent time series are based on (or derived from / related to) 
the model of multi-affine  fractional Brownian motion (mafBm), cf.\ \cite{cont,mandel3,Ivan,Amaral,Lin}. 
This is not surprising, because multifractal behaviour is believed to be the most universal case of scale-invariance, 
c.f.\ \cite{Lovejoy}. Note that while CTRW and fBm-mtt models are somewhat more detailed than a data description by 
multifractal formalism (by introducing the concept of waiting- or trading time), the time series generated by both these models
can be also characterized by the  multi-affine spectra and lead to the same scaling of low-variablity periods as the model of mafBm.

\subsection{Multi-affine  fractional Brownian motion}
One of the disadvantages of the multi-affine analysis is neglecting the long-term correlations 
in the dynamics of high-frequency variability. 
This claim is motivated as follows. Consider a multi-affine time series $x(t)$, with a lower cut-off scale $\tau_0$. 
Then,  each point of the time series $t_0$ is characterized by its own Hurst exponent $h$ (referred to as the 
Lipschitz-H\" older exponent); this exponent describes the local scaling of fluctuations,
\begin{equation}\label{mf}
|x(t)-x(t_0)| \sim |t-t_0|^h, \; \; \; |t-t_0| \gg \tau_0.
\end{equation}
Further, the distribution of points of certain values of $h$ is self-similar, and is described by a fractal 
dimension $f(h)$. So, according to the mafBm model, we study the scaling of increments  $|x(t)-x(t_0)|$ 
at $|t-t_0| \gg \tau_0$, without specific attention to the values of them at $|t-t_0| \approx \tau_0$.
In order to shed light into this phenomenon, the method of scaling analysis 
of low-variability periods has been devised \cite{kalda2}.

On the other hand, a low-variability period of length $l_i$ is defined as such a continuous time interval $T_i=[t_i,t_i+l_i]$ ($i=1,2,\ldots$) that 
{\em (a)} 
\begin{equation} \label{delta}
|x(t)-\left<x(t)\right>_\tau| \le  \delta \;\;  \mbox{for} \;\; t \in T_i
\end{equation}
where $\delta$ is a threshold parameter and angular braces denote sliding average over a window of width $\tau > \tau_0$ 
(in principle, the window width can be arbitrarily large; however, the highest time resolution of the metod 
and widest scaling range is achieved when it is as small as possible, i.e.\ just few cut-off scales $\tau_0$);
{\em (b)}
each period has maximal possible length, implying that decreasing $t_i$ or increasing $l_i$ would lead to violation of Eq.~\ref{delta}. We speak about multiscaling behaviour, if the cumulative distribution function of the low-variability periods (the number of periods with $l_i \ge n$) scales as 
\begin{equation} \label{scal}
R(n)=R_0n^{-\alpha(\delta,\tau)},
\end{equation}
where $\alpha(\delta,\tau)$ is a scaling exponent and $R_0$ is a constant.

It should be noted that the scaling exponent $\alpha(\delta,\tau)$ can be considered as the extension of the concept of multi-affine spectrum $f(h)$ to the highest possible time resolution. This is because of two circumstances.
First, the effective time resolution of the multi-affine description by 
Eq.\  \ref{mf} is limited by the required scaling range $|t-t_0| \gg \tau_0$; meanwhile, the effective time resolution of the description based on low-variability periods is the cut-off scale $\tau_0$. 
Second, in the case of infinite time resolution, there is one-to-one correspondence between the spectra $f(h)$ and $\alpha(\delta,\tau)$, see below.

The relationship between $f(h)$ and $\alpha(\delta,\tau)$ spectra can be derived as follows.
If a time moment $t$ is characterized by a Lipschitz-H\" older exponent $h$, 
the mismatch between the local average and the instant value is of the same order of 
magnitude as the variability amplitude of $x(t)$ over the time window of width $\tau$:
\begin{equation} \label{h}
|x(t)-\left<x(t)\right>_\tau|  \approx \tau^h.
\end{equation}
Comparing Eqns \ref{delta} and \ref{h}, we conclude that for a period $T_i$, all the points are characterized by $h\le \log_\tau  \delta$. Similarly, the edges $t_i-\tau_0$ and $t_i+l_i+\tau_0$ of the period $T_i$
are characterized by $h> \log_\tau  \delta$. Therefore, the the low-variability periods are the contiguous pieces which 
remain from the entire time-axis  after the removal of the high-variability points with $h> \log_\tau  \delta$.
These high-variability points form a sparse Cantor-dust-like set, assuming that $\log_\tau  \delta>h_0$, 
where $h_0$ corresponds to the global maximum of the spectrum $f(h)$ [i.e.\ $f(h_0)$=1]. 
Indeed, $f(h)$ is the fractal dimension of the set of points described by the exponent $h$. So, 
the fractal dimension of the set of points described by exponents $h>\log_\tau  \delta$ is found as
\begin{equation} \label{hh}
d=\sup_{h>\log_\tau  \delta}f(h)=\left\{
\begin{array}{ll}
1,&\mbox{~~~if~~~} \log_\tau  \delta \le h_0\\
f(\log_\tau  \delta),&\mbox{~~~if~~~} \log_\tau  \delta > h_0
\end{array}\right. .
\end{equation}
If an interval is divided into pieces by a Cantor dust of dimension $d$, 
the cumulative length-distribution exponent of the pieces is also $d$ 
(because the number of pieces larger than $n$ is estimated as the number of 
boxes of size $n$ required to 	cover the Cantor dust, according to the box-counting method). 
So, we have correspondence
\begin{equation} \label{alpha}
\alpha(\delta, \tau) = f(\log_\tau  \delta) \;\;\;\;\mbox{if}\;\;\;\; \log_\tau  \delta > h_0.
\end{equation}

Note that in the case of simple fBm (which is studied in more details in Ref.\ \cite {Carbone}), 
there is no scaling of low-variability periods with lengths $l>\tau$. Indeed,
for simple fBm, $f(h)=1$, if $h=H$ [otherwise, $f(h)$ is not defined]. 
So, the high-variability points either are absent (i.e.\ there is a single low-variability period), 
or they are populated quasi-homogeneously with fractal dimension $d=1$ (then, all the low-variability periods are shorter than
or of the order of the window width $\tau$).

\subsection{Multi-affine  trading time and discussions}
The approach used for mafBm can be also applied to the fBm-mtt model with $x(t) = F({\mathcal T}(t))$, where
$F$ is a fractional Brownian function of its argument(described by the 
Hurst exponent $H$) and  ${\mathcal T}(t)$ is a multi-affine function of time. Indeed, in that case Eq.~\ref{h} is substituted by 
\begin{equation} \label{mh}
|x(t)-\left<x(t)\right>_\tau|  \approx (\Delta_\tau T)^H \approx \tau^{hH}.
\end{equation}
Here, $\Delta_\tau T \approx\tau^h $ is the increment of the trading time ${\mathcal T}(t)$ and $h$ is the
local Lipschitz-H\" older exponent for the function ${\mathcal T}(t)$. Accordingly, Eq.~\ref{alpha} is substituted
by 
\begin{equation} \label{malpha}
\alpha(\delta, \tau) = f(H^{-1}\log_\tau  \delta) \;\;\;\;\mbox{if}\;\;\;\; \log_\tau  \delta > Hh_0.
\end{equation}

So, both in the case of the mafBm and fBm-mtt models, the scaling described by multifractal spectrum
$f(h)$ with $h>h_0$ can be also described by the scaling of low-variability periods. The approach to 
describing the Lipschitz-H\" older exponent range $h<h_0$ is straightforward: instead of low-variability
periods, high variability periods have to be used. Then, Eqns.~\ref{alpha} and \ref{malpha}
will remain valid, except that the inequalities will be opposite; details of this aspect 
will be analysed elsewhere.
Now, we are left with two questions. 

First, the multifractal dimension $f(h)$ cannot exceed the 
topological dimension one. Is it possible to have the scaling exponent of low-variability periods larger 
than one? The answer is yes (examples are provided below). More specifically, equality $\alpha > 1$ 
assumes that  the most part of the aggregated time $T_{\mbox{\scriptsize agg}}$ of the low-variability periods corresponds to 
shortest periods  (because $T_{\mbox{\scriptsize agg}} = \int n\cdot dR(n) = R_0\int n^{-\alpha-1}dn$; here, the integral 
vanishes at large lengths $n$). This assumes that the fractal dimension of the high variability periods
is one, i.e.\ $\log_\tau  \delta < h_0$. In that case, the low-variability moments form a Cantor-dust like set of points.
For real time series, the time resolution is always finite; then, in the regions of high ``dust density'', the ``dust particles''
may overlap and form relatively long (much longer than the time resolution, but much shorter than the total time interval) 
low-variability periods. Such a clustering of small-$h$-points is not described by the multifractal spectrum $f(h)$.

Second, the multifractal spectrum $f(h)$ is a one-parameter-curve; meanwhile, $\alpha(\delta,\tau)$ is a 
two-parameter-curve, which 
is effectively reduced to one degree of freedom by Eq.~\ref{alpha} [or~\ref{malpha}].
Is it possible to have a such a class of scale-invariant time series, which is more generic than the class of multi-affine functions, so that the exponent $\alpha$ can be arbitrary function of  two independent variables $\delta$ and $\tau$. This question is left open for further studies.

To conclude, the scaling analysis of low variability periods provides a simple and superior (in the sense of time resolution 
and applicability to the parameter range with $\alpha >1$) alternative to the traditional multi-affine analysis of the 
time series. In particular, Eq. \ref{hh} provides a way to check  the validity of the  assumption of multi-affinity, which requires  $\alpha(\delta, \tau) \equiv \alpha(\log_\tau  \delta)$.  If there is no need for such verification, the route is 
even simpler, e.g.\ the analysis can be performed at a fixed window length $\tau$. All this does not mean that studying 
the scaling of low-variability periods is always better than the multi-fractal formalism. Instead, the two methods 
should be considered as complementing each other, because the accuracy and applicability of both 
methods are sensitive with respect to the width of the scaling range and specifics of the scaling behaviour.

\section{Properties of low-variability periods in financial time series}
\label{powerlaw}
Now we apply the above developed theory to financial time series. The specifics of 
long-term stock-market time series is that prices can vary by orders of magnitude. Therefore,
it makes sense to work with a logarithmic scale, and substitute the definition of low variability 
(Eq.~\ref{delta}) as follows:
\begin{equation} \label{delta2}
|1-P(t)/\left<P(t)\right>_\tau| \le  \delta \;\;  \mbox{for} \;\; t \in T_i.
\end{equation}
Here, unlike in the case of Eq.~\ref{delta}, it is assumed that average is taken not over a 
time window centered around the current moment of time $t$, but instead, the large-$t$-edge of the
window coincides with $t$ (because for practical applications in real time, the future is not known):
\begin{equation}  \label{aver}
\left<P(t)\right>_{\tau}=\frac{1}{\tau}\sum_{k=0}^{\tau-1}P(t-k).
\end{equation}
So, a low-variability period is the period between two consecutive price movements that are exceeding the given threshold $\delta$. The length of a low variability period is measured in the same units as time $t$ --- in the units of the sampling period of the market data. The local average in Eqn~\ref{delta2} gives us the second input parameter: $\tau$ that could be interpreted as a degree of ``locality''. With these two input variables we maintain the easy interpretation of the results: threshold value $\delta$ is simply the percentage change of asset at time $t$ compared against the average price of that asset in preceding period with length of $\tau$ time units. 
Further we measure the lengths of low-variability periods and we count them. Then we define a cumulative distribution function $R(n)$ that represents the amounts of periods where low-variability lasted at least $n$ time units.
Finally, we fit the data against the scaling law given by  Eq.~\ref{scal}.

Following analysis will give answers to two essential questions:
\begin{enumerate}
\item Do the low variability periods obey power law and if so, what is the scaling exponent?
\item Assuming that there is a power law, how does the scaling exponent depend on the parameters $\delta$ and $\tau$?
\end{enumerate}

\subsection{Determination of Power law}
\label{power}
We used the following data in our analysis:
\begin{table}[loc=htbp]\label{algandmed}
\caption{Data used in the analysis}
\includegraphics[width=13cm]{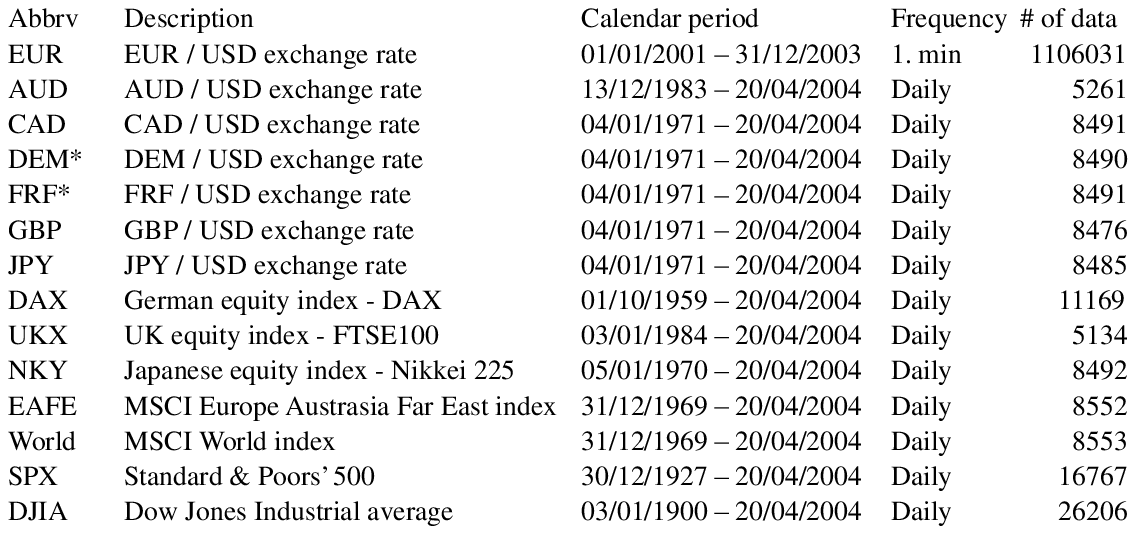}
\end{table}

*Remark: from January, 1st 1999 German Mark and French Franc are fixed against Euro.

\begin{figure}[loc=htbp]
\caption{Determination of the scaling exponent $\alpha$ for SPX time series under the following conditions: $\delta$=1.6\% and $\tau$=10 days}
\includegraphics[width=13cm]{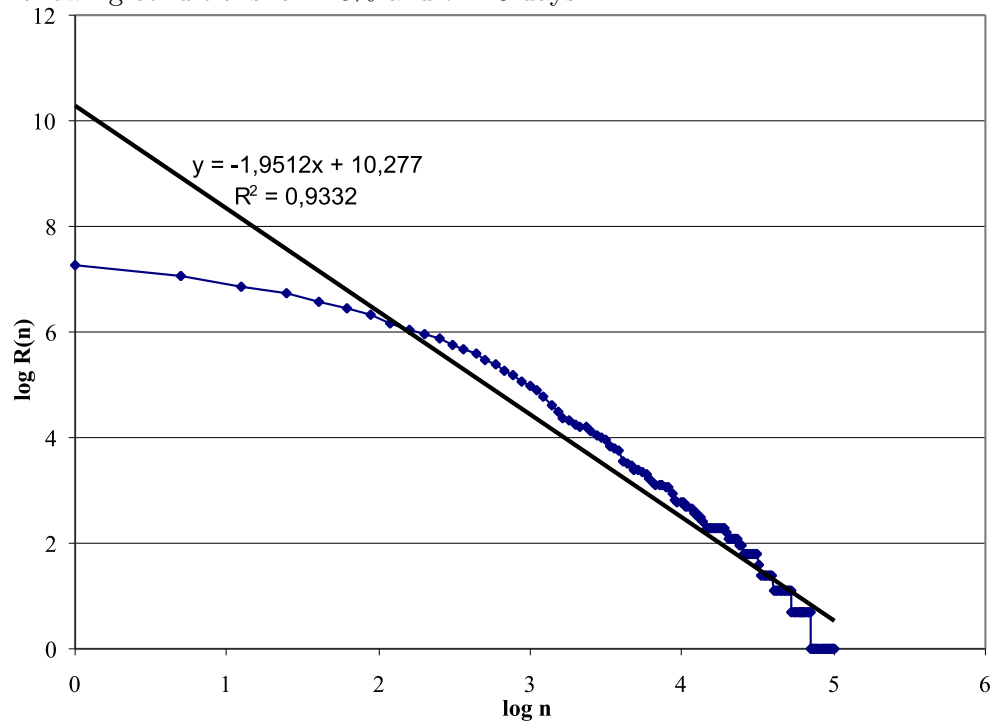}
\label{alfa}
\end{figure}

In our analysis, the scaling exponent was found by using least-squared fit for the whole data series. 
For visualizations, the dependence $R(n)$ is plotted in log-log scale. Then, the scaling exponent $\alpha$ 
is equal to slope of the fit-line.
Financial time series are single realizations of intermittently fluctuating nonstationary time series, which makes calculation of exact error estimation of the scaling exponent impossible. However, rough estimates of the uncertainties have been obtained as follows. The least-squares fitted trend-line was found as described above, except that the slope $\alpha$ was not optimized, i.e. it was considered as a fixed parameter. Further, the sum of squared residuals $r(\alpha)$ was calculated as a function of $\alpha$. The error estimate was found as $e = (\alpha'-\alpha)$, where $\alpha$ is the least-squares fitted value of the slope, and $\alpha$' satisfies the condition $r(\alpha') = 2r(\alpha)$.
The technique used for obtaining the scaling exponent $\alpha$ is illustrated in Fig~\ref{alfa}. 

\begin{figure}[loc=htbp]
\caption{Histogram of R-squared coefficients based on the regression analysis of determining scaling exponent $\alpha$}
\includegraphics[width=13cm]{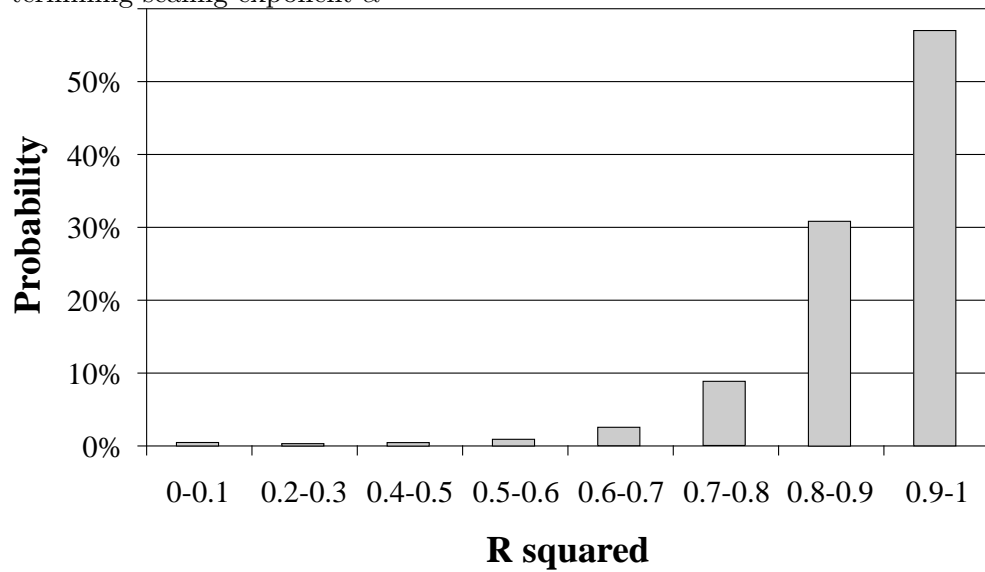}
\label{histogr}
\end{figure}

Above-mentioned procedure was carried out for all of the time series and with 
different thresholds and window widths. 
The quality of the data fit was measured by the correlation coefficient $R^2$.
Total 1004 measurements were carried out with all the daily time series. 
In Fig~\ref{histogr}, the validity of power law hypothesis is demonstrated. 
Histogram shows that in most cases, the power law provided a good data fit. 
Calculated values of $\alpha$ and error estimations for currencies and equity 
indices for $\tau = 10\,$days are presented in tables~\ref{curtable} and~\ref{eqtable} respectively.

It is important to outline two effects: 
\begin{itemize}
	\item There is reverse relationship between the correlation coefficient and the threshold parameter $\delta$. 
	The reason is that the amount of large movements is small and not representative for statistical analysis.
	\item There is also reverse relationship between the correlation coefficient and the window width $\tau$, due to similar reasons (longer averaging smoothes time series and large movements become more frequent). 
\end{itemize}
Due to these effects there is no sense to measure the scaling exponents beyond certain parameter range, e.g.\  
for $\tau=2\,$days and $\delta>2\%$. 

\begin{table}[loc=htbp]
\caption{Scaling exponents for currency time series with $\tau$=10 days}
\includegraphics[width=13cm]{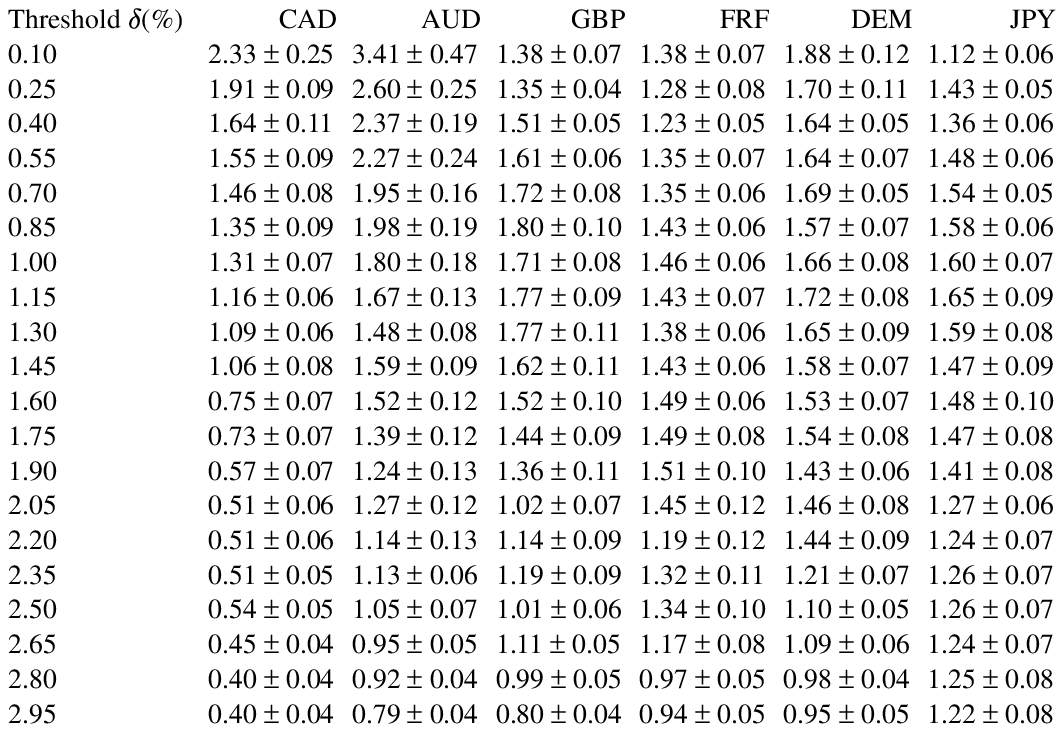}
\label{curtable}
\end{table}

\begin{table}[loc=htbp]
\caption{Scaling exponents for equity index time series with $\tau$=10 days}
\includegraphics[width=13cm]{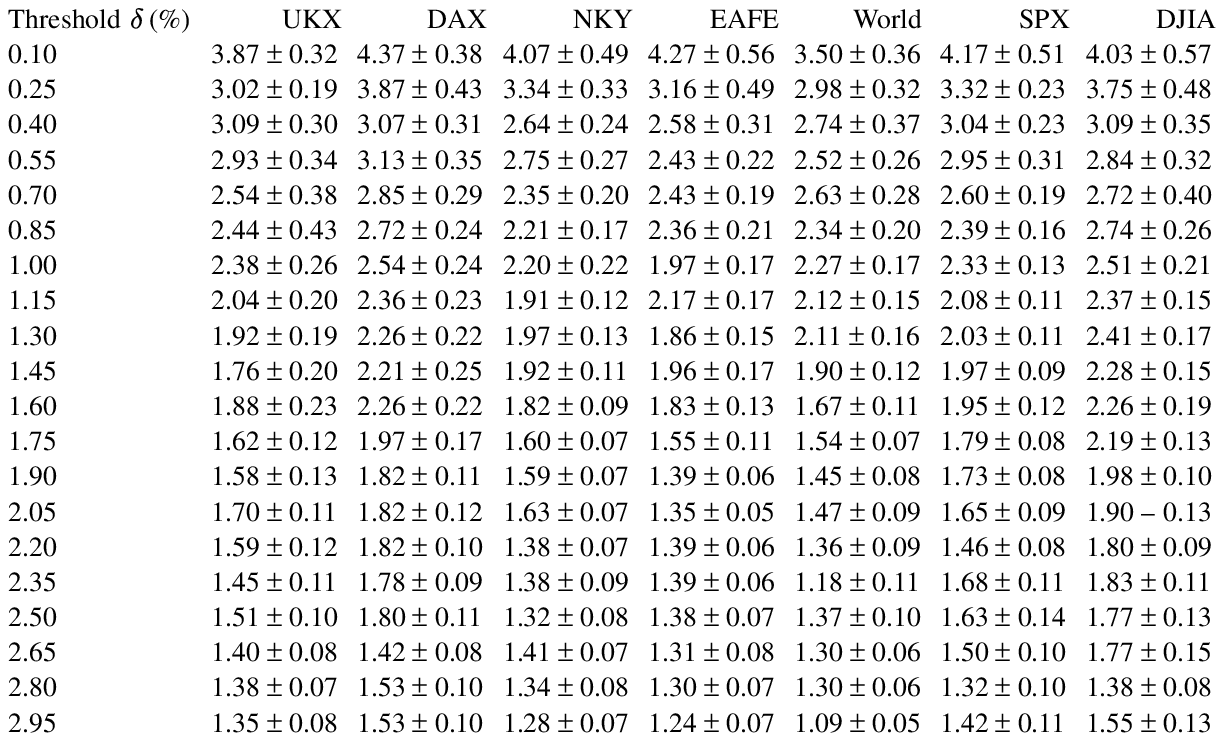}
\label{eqtable}
\end{table}

\subsection{Dependence of the scaling exponent on the parameters}\label{propalfa}

\begin{figure}[loc=htbp]
\caption{$\alpha$ values for DEM and SPX under thresholds $\delta$ = 0.10\% ...2.95\% and $\tau$=10 days.}
\includegraphics[width=13cm]{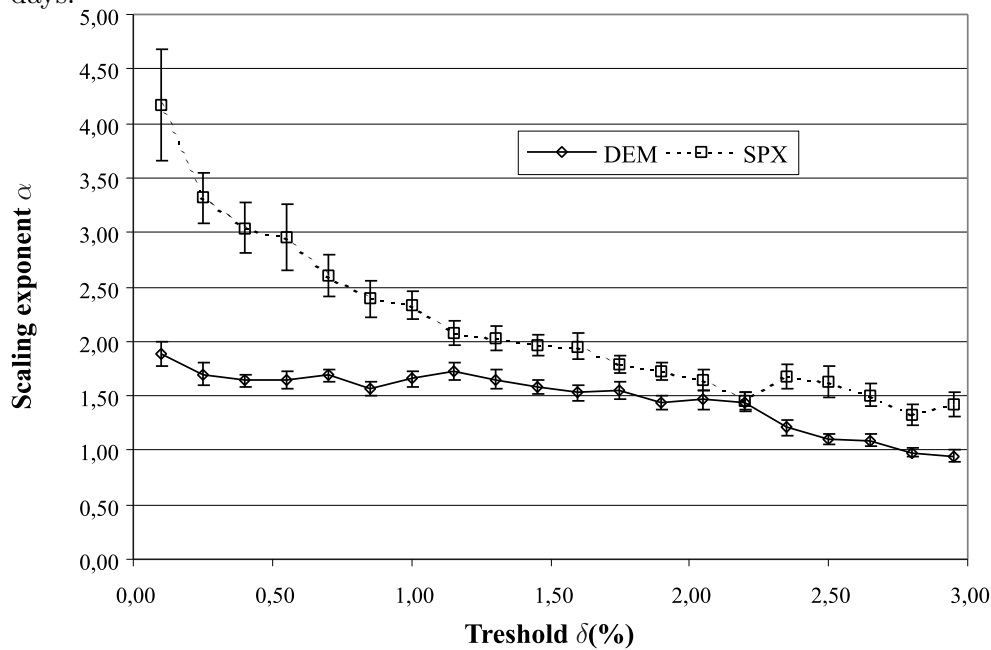}
\label{demspx}
\end{figure}

Two phenomena can be observed:
\begin{enumerate}
	\item Exponent $\alpha$ and threshold $\delta$ tend to be negatively related: larger $\delta$ values correspond to 
	lower $\alpha$ values.
	\item 	For equities, the dependence of the exponent $\alpha$ on the threshold $\delta$ is relatively 
	strong, but relatively weak for currencies. 
\end{enumerate}
In order to illustrate this observation, the values of $\alpha$ are plotted in Fig~\ref{demspx} against the threshold $\delta$ for DEM and SPX time series.

\begin{figure}[loc=htbp]
\caption{Scaling exponent $\alpha$ for EUR time series with different window widths $\tau$ under thresholds $\delta$= 0.10\%...2.95\%.}
\includegraphics[width=13cm]{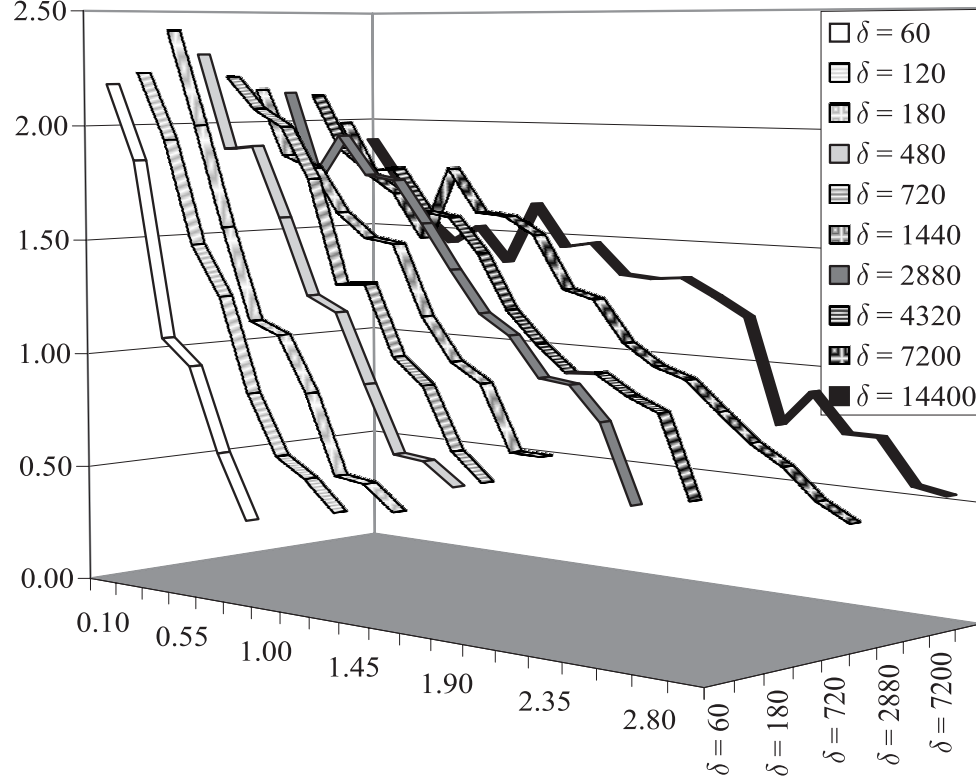}
\label{EURhifrq}
\end{figure}

Next we study the dependence of $\alpha$ on the window width $\tau$ using the high frequency data of EUR-USD exchange rate.
As mentioned above, low amount of statistics becomes an issue, when the threshold parameter is too high for the 
given window width. Therefore only those values of $\alpha$ are plotted in Fig~\ref{EURhifrq}, where $R^2>0.7$.

\subsection{Discussions of the results}
According to the theoretical expectations, $\alpha$ should be a decreasing function of $\delta$, owing to 
Eq.~\ref{alpha} and to the circumstance that for $h>h_0$ (i.e.\ for $\alpha \le 1$), 
$f(h)$ is a decreasing function of $h$.
This trend is expected to held for the range $\alpha > 1$, as well, due to simple reasoning.
Increasing the threshold parameter $\delta$ leads to some fluctuations ceasing to 
be qualified as ``large'', so that longer low-variability periods will emerge. 
Increasing the number of long periods, in its turn, corresponds to decreasing the scaling exponent $\alpha$,
in accordance with the results depicted in  Fig ~\ref{demspx}. 

\begin{figure}[loc=htbp]
\caption{Scaling exponent $\alpha$ for EUR time series with different values of $\tau$ and 
$\delta$ are plotted versus $\log_\tau\delta$, providing a data collapse for $\tau> 1\,$day, $\alpha <1$.}
\includegraphics[width=13cm]{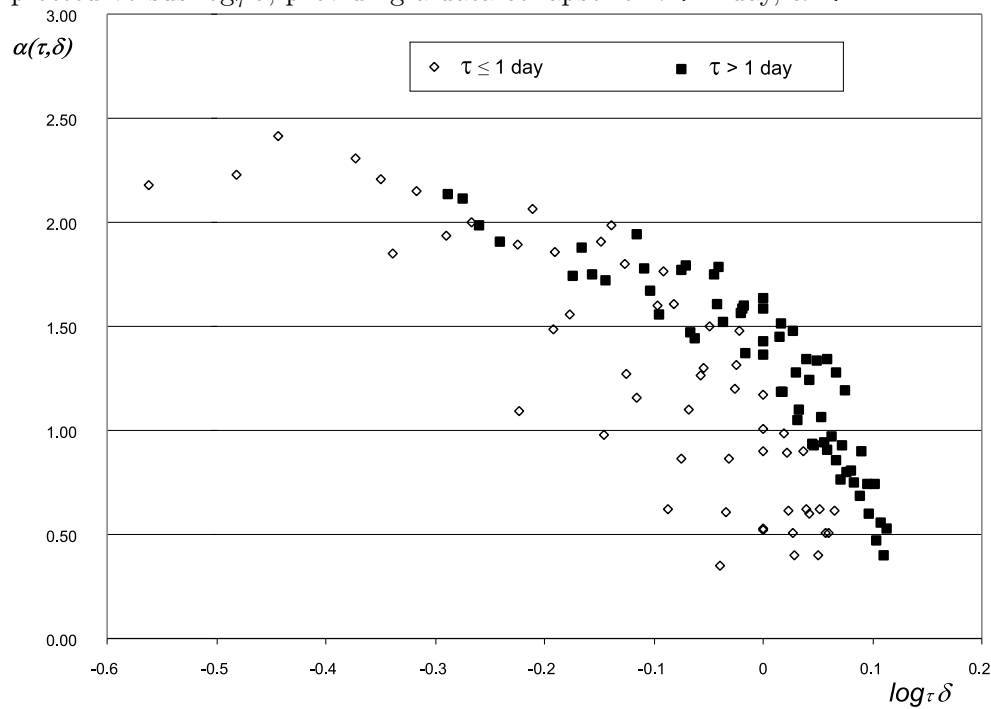}
\label{collaps}
\end{figure}

In order to study, how well the theoretically derived relationship Eq.~\ref{alpha} (or Eq.~\ref{malpha}) 
holds for the financial time series, the scaling exponent $\alpha(\delta,\tau)$ has been plotted versus
$\log_\tau\delta$ in Fig.~\ref{collaps} (using the same data as in Fig.~\ref{EURhifrq}). 
Note that the data-points with $\tau > 2880\,$min $=2\,$days describe the scaling 
behaviour at the time scale above one day; the data-points in the range $\tau \le 1\,$day 
are influenced significantly by the intra-day dynamics of the exchange rate.

For the range of validity of Eq.~\ref{alpha} ($\alpha<1$), the data-points with $\tau \ge 2$ lay 
reasonably well into a single curve, confirming ~the validity of Eq.~\ref{alpha}. 
The data of Fig.~\ref{collaps} indicate also that at the time-scale
around one day, there is a cross-over between different scaling regimes: the data with $\tau \le 1\,$day
incorporate a mixed dynamics (which cannot be described by mafBm model) 
and lay no longer on the same curve. 
At the range $\alpha > 1$, there is no evident reason for the data to lay on a curve: all the 
corresponding data points are scattered in a form of a disperse cloud.

\section{Conclusion}
In this paper a new method for time series analysis is derived and analytically motivated. 
It is shown that for multi-affine time series, the 
scaling properties of low-variability periods are described by scaling 
exponent $\alpha$ as a function of threshold parameter $\delta$ and averaging window width $\tau$.
The scaling analysis of low variability periods offers a simple 
alternative to the multi-affine analysis of the  time series, providing 
wider applicability range and somewhat higher time resolution. 
An open question is about the existence of a class of 
scale-invariant time series, more generic than multi-affine ones, 
violating Eq.~\ref{alpha}.

Particular emphasize is paid to the financial time series. 
The analysis showed a good scaling behaviour of the currency exchange rate and stock index data
and confirmed the theoretical expectations, particularly the relationship between the 
scaling exponent $\alpha$ and the model parameters $\delta$ and $\tau$ (c.f.~Eq.~\ref{alpha}).
The analysis of high-frequency data of EUR-USD exchange rate indicated that there is a cross-over between 
different scaling regimes at the time scale of one day.
It was also observed that scaling exponent values for equity time series tend to be more 
sensitive with respect to the threshold parameter than currency time series.

\section*{Acknowledgement}
The support of Estonian SF grant No.\ 5036 is acknowledged.

\end{document}